\documentclass[12pt]{article}
\usepackage{epsfig,here}
\usepackage{graphicx}
\usepackage{amsmath,amssymb}
\usepackage{latexsym}
\usepackage{bbm}

\newcommand{\term}{}
\def\K{K} 
\def\dimK{k} 

\def\Vol{\mbox{Vol}}

\def\BZ{\mathbb{Z}}
\def\BR{\mathbb{R}}

\def\CF {{\cal F}}

\def\CL {{\cal L}}

\def\CT {{\cal T}}

\def\half{\frac{1}{2}}

\newcommand{\eq}[1]{Eq.~(\ref{eq:#1})}

\newcommand{\Rep}{{\rm Rep~ }}
\newcommand{\Tr}{{\rm Tr~ }}

\newcommand{\End}{{\rm End~ }}
\newcommand{\Lie}{{\rm Lie~ }}

\def\one{{\hbox{ 1\kern-.8mm l}}}

\def\tr{{\rm tr\,}}

\begin{document}

\pagestyle{plain}

\parindent 0mm
\parskip 6pt

\vspace*{-1.8cm}

\begin{center}
{\Large\bf Compactification of superstring theory}\\[2mm] 
Michael R. Douglas\\[2mm]
\normalsize NHETC, Rutgers University, Piscataway, NJ 08855 USA\\
{\it and}
\normalsize IHES, Bures-sur-Yvette FRANCE 91440\\
E-mail: mrd@physics.rutgers.edu
\end{center} 

\vspace*{1cm} 

\section{Introduction}

Superstring theories and M theory, at present the best candidate
quantum theories which unify gravity, Yang-Mills fields and matter,
are directly formulated in ten and eleven space-time dimensions.  To
obtain a candidate theory of our four dimensional universe, one must
find a solution of one of these theories whose low energy physics is
well described by a four dimensional effective field theory (EFT),
containing the well established Standard Model of particle physics
(SM) coupled to Einstein's general relativity.
The standard paradigm for finding such solutions is
compactification, along the lines originally proposed by Kaluza and
Klein in the context of higher dimensional general relativity.
One postulates that the underlying $D$-dimensional space-time is
a product of four-dimensional Minkowski space-time, with a 
$D-4$-dimensional compact and small Riemannian manifold $\K$.
One then finds that low energy physics effectively
averages over $\K$, leading to a 
four dimensional EFT whose field content and Lagrangian are determined
in terms of the topology and geometry of $\K$.

Of the huge body of prior work on this subject, the part most relevant for
string/M theory is supergravity compactification, as in
the limit of low energies, small curvatures and weak coupling, the
various string theories and M theory reduce to ten and eleven
dimensional supergravity theories.  Many of the qualitative features
of string/M theory compactification, and a good deal of what is known
quantitatively, can be understood simply in terms of compactification
of these field theories, with the addition of a few crucial ingredients
from string/M theory.  Thus, most of this article will
restrict attention to this case, leaving many ``stringy''
topics to the articles on conformal field theory, topological string
theory and so on.  We also largely restrict attention to
compactifications based on Ricci flat compact spaces.  There is an
equally important class in which $\K$ has positive curvature; these
lead to anti-de Sitter space-times and are discussed in the articles
on AdS/CFT.

After a general review, we begin with
compactification of the heterotic string
on a three complex dimensional Calabi-Yau manifold.  This was the first
construction which led convincingly to the SM, and remains one of
the most important examples.  We then survey the various families
of compactifications to higher dimensions, with an eye on
the relations between these compactifications which follow from
superstring duality.
We then discuss some of the
phenomena which arise in the regimes of large curvature and strong
coupling.
In the final section, we bring these ideas together in a survey
of the various known four dimensional constructions.

\section{General framework}

Let us assume we are given a $D=d+\dimK$ dimensional field theory
$\CT$.  A compactification is then a $D$-dimensional space-time which
is topologically the product of a $d$-dimensional space-time with an
$\dimK$-dimensional manifold $\K$, the compactification or
``internal'' manifold, carrying a Riemannian metric and with 
definite expectation
values for all other fields in $\CT$.  These must solve the equations of
motion, and preserve $d$-dimensional Poincar\'e invariance (or, 
perhaps another $d$-dimensional symmetry group).

The most general metric ansatz for a Poincar\'e invariant
compactification is
$$
G_{IJ} = \left(\begin{matrix}
f~\eta_{\mu\nu}& 0 \\
0& G_{ij} \end{matrix}\right) ,
$$
where the tangent space indices are $0\le I<d+\dimK=D$, $0\le\mu<d$, 
and $1\le i\le \dimK$.  Here
$\eta_{\mu\nu}$ is the Minkowski metric, $G_{ij}$ is a metric on
$\K$, and $f$ is a real valued function on $\K$ called the ``warp
factor.''

As the simplest example, consider pure $D$-dimensional general relativity.
in this case, Einstein's equations reduce to Ricci flatness of $G_{IJ}$.
Given our metric ansatz, this requires $f$ to be constant, and the
metric $G_{ij}$ on $\K$ to be Ricci flat.  Thus, any $\K$ which admits
such a metric, for example the $\dimK$ dimensional torus, will lead to
a compactification.  

Typically, if a manifold admits a Ricci flat metric, it will not
be unique; rather there will be a moduli space of such metrics.
Physically, one then expects to find solutions in which the choice of
Ricci flat metric is slowly varying in $d$-dimensional space-time.
General arguments imply that such variations must be described by
variations of $d$-dimensional fields, governed by an EFT.  Given an
explicit parameterization of the family of metrics, say
$G_{ij}(\phi^\alpha)$ for some parameters $\phi^\alpha$, in principle
the EFT could be computed explicitly by promoting the parameters to
$d$-dimensional fields, substituting this parameterization into the
$D$-dimensional action, and expanding in powers of the $d$-dimensional
derivatives.  In pure GR, we would find the four-dimensional effective
Lagrangian
\begin{equation} \label{eq:EFTexample}
\CL_{EFT} = \int d^\dimK y\ \sqrt{\det G(\phi)} R^{(4)} + 
\sqrt{\det G(\phi)} G^{ik}(\phi) G^{jl}(\phi)
 \frac{\partial G_{ij}}{\partial \phi^\alpha}
 \frac{\partial G_{kl}}{\partial \phi^\beta} \ %
 \partial_\mu \phi^\alpha
 \partial_\mu \phi^\beta + \ldots .
\end{equation}

While this is easily evaluated for $\K$ a symmetric space or torus, in
general a direct computation of $\CL_{EFT}$ is impossible.  This becomes
especially clear when one learns that the Ricci flat metrics $G_{ij}$
are not explicitly known for the examples of interest.  Nevertheless,
clever indirect methods have been found that give a great deal of
information about $\CL_{EFT}$; this is much of the art of 
superstring compactification.
However, in this section, let us ignore this point and continue
as if we could do such computations explicitly.

Given a solution, one proceeds to consider its small perturbations,
which satisfy the linearized equations of motion.  If these include
exponentially growing modes (often called ``tachyons''), the solution
is unstable.\footnote{Note that this criterion is
modified for AdS compactifications.}  The
remaining perturbations can be divided into {\term massless fields},
corresponding to {\term zero modes} of the linearized equations of
motion on $\K$, and {\term massive fields}, the others.  General
results on eigenvalues of Laplacians imply that the masses of massive
fields depend on the diameter of $\K$ as
$m \sim 1/{\rm diam}(\K)$, so at energies far smaller than $m$,
they cannot be excited.\footnote{
This is not universal; given strong negative curvature on $K$, or
a rapidly varying warp factor,
one can have perturbations of small non-zero mass.}
Thus, the massive fields can be ``integrated out,'' to leave an EFT with
a finite number of fields.  In the classical approximation, this 
simply means solving their equations of motion in terms of the 
massless fields, and using these solutions to eliminate them
from the action.  At leading order in an expansion around a solution,
these fields are zero and this step is trivial; nevertheless it is useful
in making a systematic definition of the interaction terms in the EFT.

As we saw in pure GR, the configuration space parameterized by the
massless fields in the EFT, is the moduli space of compactifications
obtained by deforming the original solution.  Thus, from a
mathematical point of view, low energy EFT can be thought of as a sort
of enhancement of the concept of moduli space, and a dictionary set
up between mathematical and physical languages.  To give its next
entry, there is a natural physical metric on moduli space, defined by
restriction from the metric on the configuration space of the theory
$\CT$; this becomes the sigma model metric for the scalars in the EFT.
Because the theories $\CT$ arising from string theory are geometrically
natural, this metric is also natural from a mathematical point of
view, and one often finds that much is already known about it.
For example, the somewhat fearsome two derivative terms in
Eq. (\ref{eq:EFTexample}), are (perhaps) less so when one realizes
that this is an explicit expression for the
Weil-Peterson metric on the moduli space of Ricci flat metrics.
In any case, knowing this dictionary is essential for taking advantage
of the literature.

Another important entry in this dictionary is that the automorphism
group of a solution, translates into the gauge group in the EFT.  This
can be either continuous, leading to the gauge symmetry of Maxwell and
Yang-Mills theories, or discrete, leading to discrete gauge symmetry.
For example, if the metric on $\K$ has continuous isometry group $G$,
the resulting EFT will have gauge symmetry $G$, as in the original
example of Kaluza and Klein with $\K\cong S^1$ and $G\cong U(1)$.
Mathematically, these phenomena of ``enhanced symmetry'' are often
treated using the languages of equivariant theories (cohomology,
K-theory, etc.), stacks, and so on.

To give another example, obstructed deformations (solutions of the
linearized equations which do not correspond to elements of the
tangent space of the true moduli space) correspond to scalar fields
which, while massless, appear in the effective potential in a way
which prevents giving them expectation values.  Since the quadratic
terms $V''$ are masses, this dependence must be at cubic or higher
order.

While the preceding concepts are general and apply to compactification
of all local field theories, string and M theory add some particular
ingredients to this general recipe.
In the limits of small curvatures and weak coupling, string and M
theory are well described by the ten and eleven dimensional
supergravity theories, and thus the string/M theory
discussion usually starts with Kaluza-Klein compactification of these
theories, which we denote I, IIa, IIb, HE, HO and M.  Let us now discuss
a particular example.

\section{Calabi-Yau compactification of the heterotic string}

Contact with the SM requires finding compactifications to $d=4$ either
without supersymmetry, or with at most $N=1$ supersymmetry,
because the SM includes chiral fermions, which are incompatible with
$N>1$.  
Let us start with the $E_8\times E_8$ heterotic string
or ``HE'' theory.  This choice
is made rather than HO because only in this case can we
find the SM fermion representations as subrepresentations of the
adjoint of the gauge group.

Besides the metric,
the other bosonic fields of the HE supergravity theory are 
a scalar $\Phi$ called the dilaton, Yang-Mills
gauge potentials for the group $G\equiv E_8\times E_8$,
and a two-form gauge potential $B$ (often called the
``Neveu-Schwarz'' or ``NS'' two-form)
whose defining characteristic
is that it minimally couples to the heterotic string world-sheet.
We will need their
gauge field strengths below:  for Yang-Mills, this is a two-form
$F^a_{IJ}$ with $a$ indexing the adjoint of $\Lie G$, and for the
NS two-form this is a three-form $H_{IJK}$.
Denoting the two Majorana-Weyl spinor 
representations of $SO(1,9)$ as $S$ and $C$, 
then the fermions are
the gravitino $\psi_I \in S\otimes V$, a spin $1/2$
``dilatino' $\lambda \in C$, and the adjoint gauginos $\chi^a\in S$.
We use $\Gamma_I$ to denote Dirac matrices contracted with a ``zehnbein,''
satisfying $\{\Gamma_I,\Gamma_J\}=2 G_{IJ}$, and 
$\Gamma_{IJ}=\half [\Gamma_I,\Gamma_J]$ etc.

A local supersymmetry transformation with parameter $\epsilon$
is then
\begin{eqnarray}
\label{eq:gravsuppsi}
\delta \psi_I &=& D_I \epsilon +
  \frac{1}{8} H_{IJK} \Gamma^{JK} \epsilon \\
\label{eq:gravsupdil}
\delta \lambda &=& \partial_I\Phi \Gamma^I \epsilon
 -  \frac{1}{12} H_{IJK} \Gamma^{IJK} \epsilon \\
\label{eq:gravsupgaugino}
\delta \chi^a &=& F^a_{IJ} \Gamma^{IJ} \epsilon .
\end{eqnarray}

We now assume $N=1$ supersymmetry.
An unbroken supersymmetry is a spinor $\epsilon$ for which the
left hand side is zero, so we seek compactifications with a unique
solution of these equations.

We first discuss the case $H=0$.
Setting $\delta\psi_\mu$ in Eq. (\ref{eq:gravsuppsi}) to zero, 
we find that the warp factor $f$ must be constant.  
The vanishing of $\delta\psi_i$ 
requires $\epsilon$ to be a covariantly constant spinor.
For a six-dimensional $M$ to have a unique such
spinor, it must have $SU(3)$ holonomy, in other words $M$ must be a
Calabi-Yau manifold.  In the following we use basic facts about
their geometry.

The vanishing of $\delta\lambda$ then requires constant dilaton $\Phi$,
while the vanishing of $\delta\chi^a$
requires the gauge field strength $F$ to solve the
hermitian Yang-Mills equations,
$$
F^{2,0} = F^{0,2} = F^{1,1} = 0 .
$$
By the theorem of Donaldson and Uhlenbeck-Yau, such solutions are in
one-to-one correspondence with $\mu$-stable holomorphic vector bundles with
structure group $H$ contained in the complexification of $G$.  Choose
such a bundle $E$; by the general discussion above, the commutant of
$H$ in $G$ will be the automorphism group of the connection on $E$ and
thus the low energy gauge group of the resulting EFT.
For example, since $E_8$ has a maximal $E_6\times SU(3)$ subgroup,
if $E$ has structure group $H=SL(3)$, there is an embedding such that
the unbroken gauge symmetry is $E_6\times E_8$, realizing one of the standard
grand unified groups $E_6$ as a factor.  

The choice of $E$ is constrained by anomaly cancellation.  This
discussion (Green, Schwarz and Witten, 1987)
modifies the Bianchi identity for $H$ to
\begin{equation} \label{eq:dheqn}
dH = \tr R\wedge R - \frac{1}{30}\sum_a F^a\wedge F^a
\end{equation}
where $R$ is the matrix of curvature two-forms.
The normalization of the $F\wedge F$ term is such that
if we take $E\cong T\K$ the holomorphic tangent bundle of $\K$,
with isomorphic connection, then using the embedding we just discussed,
we obtain a solution of Eq. (\ref{eq:dheqn}) 
with $H=0$.  

Thus, we have a complete solution of the equations of motion.  General
arguments imply that supersymmetric Minkowski solutions are stable, so
the small fluctuations consist of massless and massive fields.  Let us
now discuss a few of the massless fields.  Since the EFT has $N=1$
supersymmetry, the massless scalars live in chiral multiplets, which
are local coordinates on a complex K\"ahler manifold.

First, the moduli of Ricci-flat metrics on $K$ will lead to massless
scalar fields: the
complex structure moduli, which are naturally complex, and K\"ahler
moduli, which are not.  However, in string compactification the latter
are complexified to the periods of the two-form
$B+i J$ integrated over a basis of $H_2(\K,\BZ)$, where $J$ is the
K\"ahler form and $B$ is the NS two-form.  In addition, there is
a complex field pairing the dilaton (actually, $\exp -\Phi$) and the
``model-independent axion,'' the scalar dual in $d=4$ to the two-form
$B_{\mu\nu}$.  Finally, each complex modulus of the holomorphic
bundle $E$ will lead to a chiral multiplet.
Thus, the total number
of massless uncharged chiral multiplets is 
$1+h^{1,1}(\K)+h^{2,1}(\K)+\dim H^1(\K,\End(E))$.

Massless charged matter will arise from zero modes of the gauge field
and its supersymmetric partner spinor $\chi^a$.  It is slightly easier
to discuss the spinor, and then appeal to supersymmetry to get the bosons.
Decomposing the spinors of $SO(6)$ under $SU(3)$, one obtains $(0,p)$
forms, and the Dirac equation becomes the condition that these forms
are harmonic.  By the Hodge theorem, these are in one-to-one correspondence
with classes in Dolbeault cohomology $H^{0,p}(\K,V)$, for some bundle $V$.
The bundle $V$ is obtained by decomposing the spinor into representations of
the holonomy group of $E$.  For $H=SU(3)$, the decomposition
of the adjoint under the embedding of $SU(3)\times E_6$ in $E_8$,
\begin{equation}\label{eq:decomp}
248 = (8,1) + (1,78) + (3,27) + (\bar 3,\bar 27)
\end{equation}
implies that charged matter will form ``generations'' in the $27$,
of number $\dim H^{0,1}(\K,E)$, and ``antigenerations'' in the $\bar 27$,
of number $\dim H^{0,1}(\K,\bar E) =\dim H^{0,2}(\K,E)$.  The difference
in these numbers is determined by the Atiyah-Singer index theorem to
be
$$
N_{gen} \equiv N_{27} - N_{\bar 27} = \half c_3(E) .
$$
In the special case of $E\cong T\K$, these numbers are separately
determined to be $N_{27}=b^{1,1}$ and $N_{\bar 27}=b^{2,1}$, so
their difference is $\chi(\K)/2$, half the Euler number of $K$.
In the real world, this number is $N_{gen}=3$, and matching this
under our assumptions so far is very constraining.

Substituting these zero modes into the ten-dimensional Yang-Mills
action and integrating, one can derive the $d=4$ EFT.  For example,
the cubic terms in the superpotential, usually called Yukawa couplings
after the corresponding fermion-boson interactions in the component
Lagrangian, are obtained from the cubic product of zero modes
$$
\int_\K \Omega \wedge \Tr \left( \phi_1 \wedge \phi_2 \wedge \phi_3 \right),
$$
where $\Omega$ is the holomorphic 
$\phi_i\in H^{0,1}(\K,\Rep E)$ are the zero modes, and $\Tr$ arises
from decomposing the $E_8$ cubic group invariant.  

Note the very important fact that this expression only depends on the
cohomology classes of the $\phi_i$ (and $\Omega$).  This means the
Yukawa couplings can be computed without finding the explicit harmonic
representatives, which is not possible (we don't even know the
explicit metric).  More generally, one expects to be able to
explicitly compute the superpotential and all other holomorphic
quantities in the effective Lagrangian solely from ``topological''
information (the Dolbeault cohomology ring, and its generalizations
within topological string theory).  On the other hand,
computing the K\"ahler metric in an $N=1$ EFT is usually out of reach
as it would require having explicit normalized zero modes.  Most
results for this metric come from considering closely related
compactifications with extended supersymmetry, and arguing that the
breaking to $N=1$ supersymmetry makes small corrections to this.

There are several generalizations of this construction.
First, the necessary condition to solve \eq{dheqn} 
is that the left hand side be exact, which requires
\begin{equation}\label{eq:ctwomatch}
c_2(E) = c_2(T\K) .
\end{equation}
This allows for a wide variety of $E$ to be used, so that $N_{gen}=3$
can be attained with many more $\K$.  This class of models is often 
called ``$(0,2)$ compactification'' to denote the
world-sheet supersymmetry of the heterotic string in these backgrounds.
One can also use bundles with larger structure group,
for example $H=SL(4)$ leads to unbroken $SO(10)\times E_8$, and
$H=SL(5)$ leads to unbroken $SU(5)\times E_8$.  

The subsequent breaking of the grand unified group to the Standard
Model gauge group is typically done by choosing $\K$ with non-trivial
$\pi_1$, so that it admits a flat line bundle $W$ with non-trivial
holonomy (usually called a ``Wilson line'').  One then uses
the bundle $E\otimes W$ in the above discussion, to obtain the
commutant of $H \otimes W$ as gauge group.  For example, if
$\pi_1(\K)\cong \BZ_5$, one can use $W$ whose holomony is an 
element of order $5$ in $SU(5)$, to obtain as commutant the SM gauge group
$SU(3)\times SU(2)\times U(1)$.

Another generalization is to take the three-form $H\ne 0$.  This
discussion begins by noting that for supersymmetry, we still require
the existence of a unique spinor $\epsilon$, however it will no longer
be covariantly constant in the Levi-Civita connection.  One way to
structure the problem is to note that the right hand side of
Eq. (\ref{eq:gravsuppsi}) takes the form of a connection with torsion;
the resulting equations have been discussed mathematically in
(Li and Yau, 2004).

Another recent approach to these compactifications
(Gauntlett, 2004)
starts out by arguing that $\epsilon$ cannot vanish on
$\K$, so it defines a weak $SU(3)$ structure, a local reduction of
the structure group of $T\K$ to $SU(3)$ which need not be integrable.
This structure must be present in all $N=1$,
$d=4$ supersymmetric compactifications
and there are hopes that it will lead to a useful classification 
of the possible local structures and corresponding PDE's on $\K$.

\section{Higher dimensional and extended supersymmetric compactifications}

While there are similar quasi-realistic constructions which start
from the other string theories and M theory, before we discuss these,
let us give an overview of compactifications with $N\ge 2$ supersymmetry
in four dimensions, and in higher dimensions.  These are
simpler analog models which can be understood in more depth, and their
study led to one of the most important discoveries in string/M theory, 
the theory of superstring duality.

As before, we require a covariantly constant spinor.
For Ricci flat $\K$ with other background fields zero,
this requires the holonomy of $\K$ to be one of trivial,
$SU(n)$, $Sp(n)$, 
or the exceptional holonomies $G_2$ or $Spin(7)$.
In table 1 we tabulate the possibilities
with space-time dimension $d$ greater or equal to $3$,
listing the supergravity theory, the holonomy type of
$\K$, and the type of of the resulting EFT:
dimension $d$, 
total number of real supersymmetry parameters $Ns$,
and the number of spinor supercharges $N$ (in $d=6$, 
since left and right chirality Majorana spinors are inequivalent,
there are two numbers).

\begin{table}
$$\begin{matrix}
theory& holonomy&   d&      Ns &N \\
M,II   &torus      &{\rm any} & 32 &{\rm max} \\
M      &SU(2)      &7     & 16 &1 \\
       &SU(3)      &5     &  8 &1 \\
       &G_2        &4     &  4 &1 \\
       &Sp(4)      &3     &  6 &3 \\
       &SU(4)      &3     &  4 &2 \\
       &Spin(7)    &3     &  2 &1 \\
IIa    &SU(2)      &6     & 16 &(1,1) \\
       &SU(3)      &4     &  8 &2 \\
       &G_2        &3     &  4 &2 \\
IIb    &SU(2)      &6     & 16 &(0,2) \\
       &SU(3)      &4     &  8 &2 \\
       &G_2        &3     &  4 &2 \\
HE,HO,I  &torus      &{\rm any}     & 16 &{\rm max}/2 \\
       &SU(2)      &6     &  8 &1 \\
       &SU(3)      &4     &  4 &1 \\
       &G_2        &3     &  2 &1
\end{matrix}
$$
\noindent
Table 1 -- String/M theories, holonomy groups and the resulting supersymmetry

\end{table}

\def\case#1{\subsection{#1}}

The structure of the resulting supergravity
EFT's is heavily constrained by $Ns$.
We now discuss the various possibilities.

\case{$Ns=32$} Given the supersymmetry algebra, if such a supergravity
exists, it is unique.  Thus, toroidal compactifications 
of $d=11$ supergravity, IIa and IIb supergravity lead to the same 
series of maximally supersymmetric theories.
Their structure is governed by the exceptional Lie algebra $E_{11-d}$;
the gauge charges transform in a fundamental representation of this
algebra, while the scalar fields parameterize a coset space
$G/H$ where $G$ is the maximally split real form of the Lie group  $E_{11-d}$,
and $H$ is a maximal compact subgroup of $G$.  Nonperturbative
duality symmetries lead to a further identification by a maxmimal discrete
subgroup of $G$.

\case{$Ns=16$}
This supergravity can be coupled to maximally supersymmetric
super Yang-Mills theory, which given a choice of gauge group $G$
is unique.  Thus these theories (with zero cosmological constant and
thus allowing super-Poincar\'e symmetry) are uniquely determined by the
choice of $G$.

In $d=10$, the choices $E_8\times E_8$ and $Spin(32)/\BZ_2$ which
arise in string theory, are almost uniquely determined by the 
Green-Schwarz anomaly cancellation analysis.
Compactification of these HE, HO and type I theories
on $T^n$ produces a unique theory with moduli space
\begin{equation} \label{eq:tormod}
SO(n,n+16;\BZ)\backslash SO(n,n+16;\BR)
/ SO(n,\BR)\times SO(n+16,\BR) \times \BR^+ .
\end{equation}
In KK reduction, this arises from the choice of
metric $g_{ij}$, the antisymmetric tensor $B_{ij}$ and the choice of a
flat $E_8\times E_8$ or $Spin(32)/\BZ_2$ connection on $T^n$, while a more
unified description follows from the heterotic string world-sheet
analysis.
Here the group $SO(n,n+16)$ is defined to preserve an even
self-dual quadratic form $\eta$ of signature $(n,n+16)$, for example
$\eta=(-E_8)\oplus(-E_8)\oplus I\oplus I\oplus I$ where $I$ is the form
of signature $(1,1)$ and $E_8$ is the Cartan matrix
In fact, all such forms are equivalent under orthogonal integer similarity 
transformation, so the resulting EFT is unique.
It has a rank $16+2n$ gauge group, which at generic points in moduli
space is $U(1)^{16+2n}$, but is enhanced to a non-abelian group $G$ at
special points.  To describe $G$, we first note that a point $p$ in
moduli space determines an $n$-dimensional subspace $V_p$ of
$\BR^{16+2n}$, and an orthogonal subspace $V^\perp_p$ (of varying
dimension).  Lattice points of length squared $-2$ contained in
$V^\perp_p$ then correspond to roots of the Lie algebra of $G_p$.

The other compactifications with $Ns=16$ is M theory on K3 and its
further toroidal reductions, and IIb on K3.  M theory compactification
to $d=7$ is dual to heterotic on $T^3$, with the same moduli space
and enhanced gauge symmetry.  As we discuss at the end of section 5,
the extra massless gauge
bosons of enhanced gauge symmetry are M2 branes wrapped on two-cycles
with topology $S^2$.  For such a cycle to have zero volume, the integral
of the K\"ahler form and holomorphic two-form over the cycle must vanish;
expressing this in a basis for $H^2(K3,\BR)$ leads to
exactly the same condition we discussed for enhanced gauge symmetry
above.  The final result is that all such K3 degenerations lead to 
one of the two dimensional canonical singularities, of types A, D or
E, and the corresponding EFT phenomenon is the enhanced gauge symmetry
of corresponding Dynkin type A, D, or E.

IIb on K3 is similar, but reducing the self-dual RR four-form
potential on the two-cycles leads to self-dual tensor multiplets
instead of Maxwell theory.  The moduli space is \eq{tormod} but with
$n=5$, not $n=4$, incorporating periods of RR potentials and the
$SL(2,\BZ)$ duality symmetry of IIb theory.

One may ask if the $Ns=16$ I/HE/HO theories in $d=8$ and $d=9$
have similar duals.  For $d=8$, these are obtained by
a pretty construction known as ``F theory.''
Geometrically, the simplest definition of F theory
is to consider the special case of M theory on an elliptically fibered
Calabi-Yau, in the limit that the K\"ahler modulus of the fiber becomes
small.  One check of this claim for $d=8$ is that
the moduli space of elliptically fibered K3's agrees with
Eq. (\ref{eq:tormod}) with $n=2$.

Another definition of F theory is the particular case of IIb
compactification using Dirichlet seven-branes, and orientifold
seven-planes.  This construction is T-dual to the type I theory on
$T^2$, which provides its simplest string theory definition.  As
discussed in (Polchinski, 1999), one can think of the open strings
giving rise to type I gauge symmetry as living on $32$ Dirichlet
nine-branes (or D$9$-branes) and an orientifold nineplane.
T-duality converts Dirichlet and orientifold $p$-branes to $p-1$-branes;
thus this relation follows by applying two T-dualities.

These compactifications can also be parameterized by elliptically
fibered Calabi-Yaus, where $\K$ is the base, and the branes
correspond to singularities of the fibration.  The relation between
these two definitions follows fairly simply from the
duality between M theory on $T^2$, and IIb string on $S^1$.
There is a partially understood generalization of this to $d=9$.

Finally, these constructions admit further discrete choices, which
break some of the gauge symmetry.  The simplest to explain is in the
toroidal compactification of I/HE/HO.  The moduli space of theories we
discussed uses flat connections on the torus which are continuously
connected to the trivial connection, but in general the moduli
space of flat connections has other components.  The simplest example
is the moduli space of flat $E_8\times E_8$ connections on $S^1$, which
has a second component in which the holonomy exchanges the two $E_8$'s.
On $T^3$, there are connections for which the holonomies
cannot be simultaneously diagonalized.  This structure
and the M theory dual of these choices is discussed in (de Boer et al, 2001).

\case{$Ns=8$, $d<6$}
Again, the gravity multiplet is uniquely determined, so the most basic
classification is by the gauge group $G$.  The full low energy EFT is 
determined by the matter content and action, and
there are two types of matter multiplet.

First, vector multiplets contain
the Yang-Mills fields, fermions and $6-d$ scalars; their action is 
determined by a prepotential which is a $G$-invariant function of the 
fields.
Since the vector
multiplets contain massless adjoint scalars, a generic vacuum in which
these take non-zero distinct VEV's will have $U(1)^r$ gauge symmetry,
the commutant of $G$ with a generic matrix (for $d<5$, while there
are several real scalars, the potential forces these to commute in
a supersymmetric vacuum).  Vacua with this type of gauge symmetry
breaking, which does not reduce the rank of the gauge group,
are usually referred to as on a ``Coulomb branch'' of the
moduli space.
To summarize, this sector can be specified by $n_V$,
the number of vector multiplets, and the prepotential $\CF$, a 
function of the $n_V$ VEV's which is
cubic in $d=5$, and holomorphic in $d=4$.

Hypermultiplets contain scalars which parameterize a quaternionic
K\"ahler manifold, and partner fermions.  
Thus, this sector is specified by 
a $4n_H$ real dimensional 
quaternionic K\"ahler manifold. 
The $G$ action comes with triholomorphic moment maps; if nontrivial,
VEVs in this sector can break gauge symmetry and reduce it in rank.
Such vacua are usually referred to as on a ``Higgs branch.''

The basic example of these compactifications
is M theory on a Calabi-Yau threefold.
Reduction of the three-form leads to $h^{1,1}(\K)$ vector multiplets, whose
scalar components are the CY
K\"ahler moduli.  The CY complex structure moduli pair with periods of
the three-form to produce $h^{2,1}(\K)$ hypermultiplets.  
Enhanced gauge symmetry then appears when the CY$_3$ contains ADE
singularities fibered over a curve, from the same mechanism involving
wrapped M2 branes
we discussed
under $Ns=16$.  If degenerating curves lead to other singularities
(for example the ODP or ``conifold''), it is possible to obtain
extremal transitions which translate physically into
Coulomb-Higgs transitions.  Finally, singularities in which
surfaces degenerate lead to non-trivial fixed point theories.

Reduction on $S^1$ leads to IIa on CY$_3$, with the spectrum above
plus a ``universal hypermultiplet'' which includes the dilaton.
Perhaps the most interesting new feature is the presence of world-sheet
instantons, which correct the metric on vector multiplet moduli space.
This metric satisfies the restrictions of special geometry
and thus can be derived from a prepotential.  

The same theory can be obtained by compactification of IIb theory on
the mirror CY$_3$.  Now vector multiplets are related to the
complex structure moduli space, while hypermultiplets are related to
K\"ahler moduli space.  In this case, the prepotential derived from
variation of complex structure receives no instanton
corrections, as we discuss in the next section.

Finally, one can compactify the heterotic string on $K3\times
T^{6-d}$, but this theory follows from toroidal
reduction of the $d=6$ case we discuss next.

\case{$Ns=8$, $d=6$}
These supergravities are similar to $d<6$, 
but there is a new type of matter multiplet,
the self-dual tensor (in $d<6$ this is dual to
a vector multiplet).  Since fermions in $d=6$ are chiral,
there is an anomaly cancellation condition relating the numbers of the
three types of multiplets 
(Aspinwall, 1996, section 6.6),
\begin{equation} \label{eq:sixanom}
n_H - n_V + 29 n_T = 273.
\end{equation}

One class of examples is
the heterotic string compactified on K3.  In the original
perturbative constructions, to satisfy
Eq. (\ref{eq:ctwomatch}), 
we need to choose a vector bundle
with $c_2(V)=\chi(K3)=24$.  The resulting
degrees of freedom are a
a single self-dual tensor multiplet and a rank $16$ gauge group.
More generally, one can introduce $N_{5B}$
heterotic fivebranes, which
generalize Eq. (\ref{eq:ctwomatch}) to $c_2(E) + N_{5B} = c_2(T\K)$.
Since this brane carries a self-dual tensor multiplet, this series
of models is parameterized by $n_T$.  They are connected by transitions
in which an $E_8$ instanton shrinks to zero size and becomes a fivebrane;
the resulting decrease in the dimension of the moduli space of
$E_8$ bundles on K3 agrees with Eq. (\ref{eq:sixanom}).

Another class of examples is F theory on an elliptically
fibered Calabi-Yau threefold.  These are related to M theory on
an elliptically fibered CY$_3$ in the same general way we discussed
under $Ns=16$.  
The relation between F theory and the heterotic string
on K3 can be seen by lifting M theory-heterotic duality; this
suggests that the two constructions are dual only if the CY$_3$ is
a K3 fibration as well.  Since not all elliptically fibered CY$_3$'s
are K3 fibered, the F theory construction is more general.


We return to $d=4$ and $Ns=4$ in the final section.
The cases of
$Ns< 4$ which exist in $d\le 3$ are far less
studied.

\section{Stringy and quantum corrections}

The $D$-dimensional low energy effective supergravity actions on which
we based our discussion so far are only approximations to the general
story of string/M theory compactification.  However, if Planck's constant
is small, $K$ is sufficiently large, and its curvature is small, 
they are controlled approximations.  

In M theory, as in any theory of quantum gravity, corrections are
controlled by the Planck scale parameter $M_P^{D-2}$, which sits in
front of the Einstein term of the $D$-dimensional effective
Lagrangian, and plays the role of $\hbar$.  In general, this is
different from the four dimensional Planck scale, which satisfies
$M_{Planck\ 4}^2=\Vol(K)M_P^{D-2}$.  After taking the low energy limit
$E << M_P$, the remaining corrections are controlled by the
dimensionless parameters $l_P/R$ where $R$ can any characteristic
length scale of the solution: a curvature radius, the length of a
non-trivial cycle, and so on.

In string theory, one usually thinks of the corrections as a double
series expansion in $g_s$, the dimensionless (closed) string coupling
constant, and $\alpha'$, the inverse string tension parameter, of
dimensions $({\rm length})^2$.  The ten dimensional Planck scale is
related to these parameters as $M_P^8 = 1/g_s^2 (\alpha')^4$, up to a
constant factor which depends on conventions.  

Besides perturbative corrections, which have power-like dependence on
these parameters, there can be world-sheet and ``brane'' instanton
corrections.  For example, a string-world sheet can wrap around a
topologically non-trivial space-like two-cycle $\Sigma$ in $\K$,
leading to an instanton correction to the effective action which is
suppressed as $\exp -\Vol(\Sigma)/2\pi\alpha'$.  More generally, any
$p$-brane wrapping a $p$-cycle can produce a similar effect.  
As for which terms in the effective Lagrangian receive
corrections, this depends largely on the number and symmetries of the
fermion zero modes on the instanton world-volumes.

Let us start by discussing some cases in which one can argue that
these corrections are not present.  First, extended supersymmetry can
serve to eliminate many corrections.  This is analogous to the
familiar result that the superpotential in $d=4$, $N=1$ supersymmetric
field theory does not receive (or ``is protected from'') perturbative
corrections, and in many cases follows from similar formal arguments.
In particular, supersymmetry forbids corrections to the potential and
two derivative terms in the the $Ns=32$ and $Ns=16$ lagrangians.

In $Ns=8$, the superpotential is protected, but the two
derivative terms can receive corrections.  However, there is a simple
argument which precludes many such corrections -- since vector
multiplet and hypermultiplet moduli spaces are decoupled, a correction
whose control parameter sits in (say) a vector multiplet, cannot
affect hypermultiplet moduli space.  
This fact allows for
many exact computations in these theories.

As an example, in IIb on CY$_3$, the metric on vector multiplet moduli
space is precisely \eq{EFTexample} as obtained from supergravity, in
other words the Weil-Peterson metric on complex structure moduli
space.  First, while in principle it could have been corrected by
world-sheet instantons, since these depend on K\"ahler moduli which
sit in hypermultiplets, it is not.  The only other instantons with the
requisite zero modes to modify this metric are wrapped Dirichlet
branes.  Since in IIb theory these wrap even dimensional cycles, they
also depend on K\"ahler moduli and thus leave vector moduli space
unaffected.

As previously discussed, for K3-fibered CY$_3$,
this theory is dual to the heterotic string on $K3\times T^2$.
There, the vector multiplets arise from Wilson lines on $T^2$, and
reduce to an adjoint multiplet of $N=2$ supersymmetric Yang-Mills theory.
Of course, in the quantum theory,
the metric on this 
moduli space receives instanton corrections.  Thus, the duality
allows deriving the exact moduli space metric, and
many other results of the Seiberg-Witten theory of $N=2$
gauge theory, as aspects of the geometry of Calabi-Yau moduli space.

In $Ns=4$, only the superpotential is protected, and that only in
perturbation theory; it can receive non-perturbative corrections.
Indeed, it appears that this is fairly generic, suggesting that the
effective potentials in these theories are often sufficiently
complicated to exhibit the structure required for supersymmetry
breaking and the other symmetry breakings of the SM.  Understanding
this is an active subject of research.

We now turn from corrections to novel physical phenomena which arise
in these regimes.  While this is too large a subject to survey here,
one of the basic principles which governs this subject is the idea
that string/M theory compactification on a singular manifold $\K$ is
typically consistent, but has new light degrees of freedom in the EFT,
not predicted by Kaluza-Klein arguments.  We implicitly touched on one
example of this in the discussion of M theory compactification on K3
above, as the space of Ricci-flat K3 metrics has degeneration
limits in which curvatures grow without bound, while the volumes of
two-cycles vanish.  On the other hand, the structure of $Ns=16$
supersymmetry essentially forces the $d=7$ EFT in these limits to be
non-singular.  Its only
noteworthy feature is that a non-abelian gauge symmetry
is restored, and thus certain charged vector bosons and their
superpartners become massless.

To see what is happening microscopically, we must consider
an M theory membrane (or two-brane), wrapped on a degenerating
two-cycle.
This appears as a particle in $d=7$,
charged under the vector potential obtained by reduction of the 
$D=11$ three-form potential.  The mass of this particle is the volume
of the two-cycle multiplied by the membrane tension, so as this volume
shrinks to zero, the particle becomes massless.  Thus the physics is
also well-defined in eleven dimensions, though not literally described
by eleven dimensional supergravity.

This phenomenon has numerous generalizations.  Their common point is
that, since the essential physics involves new light degrees of
freedom, they can be understood in terms of a lower dimensional
quantum theory associated with the region around the singularity.
Depending on the geometry of the singularity, this is sometimes a
weakly coupled field theory, and sometimes a non-trivial conformal field
theory.  Occasionally, as in IIb on K3, the lightest wrapped brane is
a string, leading to a ``little string theory'' (Aharony, 2000)

\section{$N=1$ supersymmetry in four dimensions}

Having described the general framework, we conclude by discussing the various
constructions which lead to $N=1$ supersymmetry.
Besides the heterotic string on a CY$_3$, these compactifications
include type IIa and IIb on orientifolds of Calabi-Yau threefolds, the
related F theory on elliptically fibered Calabi-Yau fourfolds, and M
theory on $G_2$ manifolds.  Let us briefly spell out their
ingredients, the known non-perturbative corrections to the superpotential,
and the duality relations between these constructions.

To start, we recap the heterotic string construction.  We must specify
a CY$_3$ $\K$, and a bundle $E$ over $\K$ which admits a hermitian Yang-Mills
connection.  The gauge group $G$ is the commutant of the structure group
of $E$ in $E_8\times E_8$ or $Spin(32)/\BZ_2$, while the chiral matter
consists of metric moduli of $\K$, and fields corresponding to a basis for
the Dolbeault cohomology group $H^{0,1}(\K,\Rep E)$
where $\Rep E$ is the bundle $E$ embedded into an $E_8$ bundle and 
decomposed into $G$-reps.

There is a general (though somewhat formal) expression for the
superpotential,
\begin{equation} \label{eq:W}
W = \int \Omega \wedge 
 + \Tr \left(\bar A \bar \partial \bar A + \frac{2}{3}\bar A^3\right) +
+ \int \Omega \wedge H^{(3)}
+ W_{NP} .
\end{equation}
The first term is the holomorphic 
Chern-Simons action, whose variation enforces the $F^{0,2}=0$
condition.  The second is the ``flux superpotential,'' while the
third term is the non-perturbative corrections.  The best understood
of these arise from supersymmetric gauge theory sectors.
In some but not all cases, these can {be understood as arising from
gauge theoretic instantons, which can be shown to be
dual to heterotic five-branes wrapped on $\K$.
Heterotic world-sheet instantons can also contribute.

The HO theory is S-dual to the type I string,
with the same gauge group, realized by open strings on Dirichlet nine-branes.
This construction involves essentially the same data.  The two classes
of heterotic instantons are dual to D$1$ and D$5$-brane instantons, whose
world-sheet theories are somewhat simpler.

If the CY$_3$ $\K$ has a fibration by tori, by applying T-duality to
the fibers along the lines discussed for tori under $Ns=16$ above,
one obtains various type II orientifold compactifications.
On an elliptic
fibration, double T-duality produces a IIb compactification with 
D$7$'s and O$7$'s.  Using the relation between IIb theory on $T^2$ and
F theory on K3 fiberwise, one can also think of this as an F theory
compactification on a K3-fibered Calabi-Yau fourfold.  More
generally, one can compactify F theory on any elliptically fibered
fourfold to obtain $N=1$.  These theories have D$3$-instantons, the
T-duals of both the type I D$1$ and D$5$-instantons.

The theory of mirror symmetry predicts that all CY$_3$'s have $T^3$
fibration structures.  Applying the corresponding triple T duality,
one obtains a IIa compactification on the mirror CY$_3$ $\tilde \K$,
with D$6$-branes and O$6$-planes.  Supersymmetry requires these to
wrap special Lagrangian cycles in $\tilde \K$.  As in all Dirichlet brane 
constructions, enhanced gauge symmetry arises from coincident branes
wrapping the same cycle, and only the classical groups are visible in
perturbation theory.  Exceptional gauge symmetry arises as a
strong coupling phenomenon of the sort described in section 5.
The superpotential can
also be thought of as mirror to Eq. (\ref{eq:W}), but now the first
term is the sum of a real Chern-Simons action on the special
Lagrangian cycles, with disk world-sheet instanton corrections, as
studied in open string mirror symmetry.  The gauge theory instantons
are now D$2$-branes.

Using the duality relation between the IIa string and eleven
dimensional M theory, this construction can be lifted to a
compactification of M theory on a seven dimensional manifold $L$,
which is an $S^1$ fibration over $\tilde \K$.
The D$6$ and O$6$ planes arise from singularities in the $S^1$
fibration.  Generically, $L$ can be smooth, and the only candidate
in table I for such an $N=1$ compactification is a manifold with
$G_2$ holonomy; therefore $L$ must have such holonomy.  Finally,
both the IIa world-sheet instantons and the D$2$-instantons 
lift to membrane instantons in M theory.

This construction implicitly demonstrates the existence of a large
number of $G_2$ holonomy manifolds.  Another way to arrive at these is
to go back to the heterotic string on $\K$, and apply the duality
(again under $Ns=16$) between heterotic on $T^3$ and M theory on K3
to the $T^3$ fibration structure on $\K$, to arrive at M theory
on a K3-fibered manifold of $G_2$ holonomy.  Wrapping membranes on
two-cycles in these fibers, we can see enhanced gauge symmetry in
this picture fairly directly.  It is an illuminating exercise to
work through its dual realizations in all of these constructions.

Our final construction uses the interpretation of the strong coupling
limit of the HE theory as M theory on a
one-dimensional interval $I$, in which the two $E_8$ factors live on
the two boundaries.  Thus, our original starting point can also be
interpreted as the heterotic string on $\K\times I$. This construction
is believed to be important physically as it allows generalizing a
heterotic string tree-level relation between the gauge and
gravitational couplings which is phenomenologically disfavored.  One
can relate it to a IIa orientifold as well, now with D$8$ and O$8$-branes.

These multiple relations are often referred to as the ``web'' of
dualities.  They lead to numerous relations between compactification
manifolds, moduli spaces, superpotentials, and other properties of the
EFT's, whose full power has only begun to be appreciated.\\

\section*{Further reading}

Original references for all but the most recent of these topics can
be found in the following textbooks and proceedings.  We have
also referenced a few research articles which are good starting points
for the more recent literature.  There are far more reviews than we could
reference here, and a partial listing of these appears at
{\tt http://www.slac.stanford.edu/spires/reviews/}

{\it Superstring Theory},
M.~B.~Green, J.~H.~Schwarz and E.~Witten, 
2 vols, Cambridge University Press, 1987.

{\it Les Houches 1995: Quantum symmetries},
eds. A. Connes and K. Gaw\c{e}dzki, 
North Holland, 1998.

{\it String Theory}, J.~Polchinski, 
2 vols,
Cambridge University Press, 1998.

{\it Quantum fields and strings: a course for mathematicians}, 
eds. P. Deligne et\ al.,
American Mathematical Society, 1999.

{\it Les Houches 2001: Unity from Duality: Gravity, Gauge theory and Strings},
eds. C. Bachas  et\ al.,
Springer 2002.

{\it Strings and Geometry: proceedings of the 2002 Clay School},
eds. M. Douglas et\ al.,
American Mathematical Society, 2004.

O.~ Aharony,
A brief review of 'little string theories',
Class.Quant.Grav.17:929-938, 2000.

P.~S.~Aspinwall,
K3 surfaces and string duality,
1996 preprint,  arXiv:hep-th/9611137.

J.~de Boer et al.,
Triples, fluxes, and strings,
Adv.\ Theor.\ Math.\ Phys.\  {\bf 4}, 995 (2002).

J. Gauntlett,
Branes, calibrations and supergravity,
in Strings and Geometry, Douglas et al., pp. 79-126, AMS 2004.

J. Li and S.-T. Yau, 
The existence of supersymmetric string theory with torsion,
2004 preprint, arXiv:hep-th/0411136

\end{document}